\documentclass[iop]{emulateapj}

\usepackage[varg]{txfonts}
\usepackage{enumerate}
\usepackage{enumitem}
\usepackage{float}
\usepackage{array}
\usepackage[english]{babel}

\usepackage{amsmath}

\usepackage{natbib}

\defcitealias{wahlberg14}{WJJ}
\defcitealias{bukhari16}{Paper I}

\newcommand{\fref}[1]{Fig. \ref{#1}}			
\newcommand{\frefp}[1]{(Fig. \ref{#1})}			
\newcommand{\sref}[1]{Section \ref{#1}}			
\newcommand{\tref}[1]{Table \ref{#1}}			

\newcommand{\eref}[1]{Eq. \eqref{#1}}			
\newcommand{\erefnp}[1]{Eq. \ref{#1}}			
\newcommand{\erefp}[1]{(Eq. \ref{#1})}			
\newcommand{\erefCon}[2]{Eqs. \eqref{#1}--\eqref{#2}}	

\let\t=\tsub                 
\newcommand{\t}[1]{\text{\textnormal{#1}}}  

\makeatletter
\newcommand{\thickhline}{%
    \noalign {\ifnum 0=`}\fi \hrule height 1.2pt
    \futurelet \reserved@a \@xhline
}
\newcolumntype{"}{@{\hskip\tabcolsep\vrule width 1.2pt\hskip\tabcolsep}}
\makeatother

\begin{document}

\title{The role of pebble fragmentation in planetesimal formation\\II. Numerical simulations}

\author{Karl Wahlberg Jansson\altaffilmark{1}, Anders Johansen\altaffilmark{1}, Mohtashim Bukhari Syed\altaffilmark{2}, and J\"urgen Blum\altaffilmark{2}}
\affil{$^1$\ Lund Observatory, Department of Astronomy and Theoretical Physics, Lund University, Box 43, SE-221 00 Lund, Sweden\\ $^2$\ Technische Universit\"at Braunschweig, Institut f\"ur Geophysik und extraterrestrische Physik, Mendelssohnstra\ss e 3, D-38106 Braunschweig, Germany}

\begin{abstract}

\noindent Some scenarios for planetesimal formation go through a phase of collapse of gravitationally bound clouds of mm-cm-sized pebbles. Such clouds can form for example through the streaming instability in protoplanetary disks. We model the collapse process with a statistical model to obtain the internal structure of planetesimals with solid radii between 10 and 1,000 km. In the collapse, pebbles collide and, depending on relative speed, collisions have different outcomes. A mixture of particle sizes inside a planetesimal leads to better packing capabilities and higher densities. In this paper we apply results from new laboratory experiments of dust aggregate collisions (presented in a companion paper) to model collision outcomes. We find that the internal structure of a planetesimal is strongly dependent on both its mass and the applied fragmentation model. Low-mass planetesimals have no/few fragmenting pebble collisions in the collapse phase and end up as porous pebble-piles. The amount of fragmenting collisions increases with increasing cloud mass, resulting in wider particle size distributions and higher density. The collapse is nevertheless ``cold'' in the sense that collision speeds are damped by the high collision frequency. This ensures that a significant fraction of large pebbles survive the collapse in all but the most massive clouds. Our results are in broad agreement with the observed increase in density of Kuiper belt objects with increasing size as exemplified by the recent characterization of the highly porous comet 67P/Churyumov-Gerasimenko.

\end{abstract}

\keywords{Methods: analytical -- Methods: numerical -- Planets and satellites: formation}

\maketitle

\section{Introduction}\label{sec:intro}

Planet formation takes place around young stars as $\mu$m-sized dust and ice particles grow to ever larger bodies \citep{safronov69}. This leads to planets of characteristic sizes $10^4$-$10^5$ km. The growth starts with particles sticking together by contact forces \citep[see review by][]{blum08}. Compactified pebbles of mm-cm sizes have poor sticking properties, but growth to planetesimals can be aided by the mutual gravity in pebble clouds that are concentrated in turbulent gas \citep[see review by][]{johansen14}. This leads to the formation of planetesimals with a distribution of sizes ranging from 10 to several 100 km \citep{johansen15,simon15}.

Many details of the gravitational collapse phase are not yet fully understood. \citet{nesvorny10} pioneered the modelling of the collapse phase in N-body simulations of a large number of pebbles coming together by their mutual gravity. They found that pebble clouds with high internal angular momentum collapse into binary planetesimals. This can explain the high fraction of binaries observed in the classical cold population of trans-Neptunian objects \citep{noll08}. The two components in binary Kuiper belt objects appear to have the same colour and composition \citep{benecchi09}, suggesting that they formed together, since the Kuiper belt, overall, has a broad colour distribution. In the case of binary formation through three-body encounters \citep[e.g.][]{goldreich02} the components, most likely, should not have the same composition. 

\medskip
\citet[hereafter WJJ]{wahlberg14} investigated the evolution of the particle size distribution during the collapse phase, based on laboratory experiments of particle collisions \citep{guttler10}. The initial pebble clouds were assumed to arise from the streaming instability \citep[e.g.][]{youdin05,johansen09,johansen12,bai10}. A major result of this paper was that the collapse process is very dependent on planetesimal mass. More massive clouds collapse faster and collisions between pebbles result in pebble fragmentation. This affects the internal structure (e.g. density and porosity) of the resulting planetesimal: low-mass planetesimals should be porous pebble-piles while higher-mass planetesimals are a denser mixture of dust and pebbles. This relation with density increasing with increasing size agrees with observations of Kuiper belt objects \citep{brown13}. However, other parameters such as composition, radioactive heating and collisions will also affect the structure, but the effect of those processes all depend on the initial porosity and packing efficiency. Therefore the outcome of pebble cloud collapse models can be used as starting point for calculations of the long-term thermal evolution of planetesimals \citep[e.g.][]{lichtenberg16}. \citet{lorek16} expanded the cloud collapse model in \citetalias{wahlberg14} to investigate the evolution of the density of the pebbles throughout the collapse. Bouncing collisions cause porous dust (ice) aggregates to become more compact. The authors used their results to constrain the range of initial conditions (cloud mass, dust-to-ice ratio and initial filling factor) that can produce comets and other observed bodies in the outer Solar System. They find that planetesimals with observed comet bulk density of $\sim$0.5 g cm$^{-3}$ can form either if the cloud is low-mass and initially have compact pebbles or if the cloud is massive independent on initial pebble porosity.

\medskip
Observational data on the structure of planetesimals in the outer Solar System has increased dramatically in the past years. The space probes Rosetta (ESA) and New Horizons (NASA) have both reached their respective targets, the Jupiter family comet 67P/Churyumov-Gerasimenko (hereafter 67P) and the dwarf planet Pluto.

Rosetta has multiple instruments that provide measurements for understanding the origin of planetesimals. OSIRIS is an optical, spectroscopic and infrared system for imaging the nucleus of 67P from Rosetta \citep{keller07}. The ``goosebump'' structures in the walls of the deep pits have been suggested to represent the primordial pebbles that make up the bulk of the comet \citep{sierks15}, although the meter scale of those pebbles are in some disagreement with the particle sizes that are believed possible to form by coagulation in the outer regions of protoplanetary disks \citep{birnstiel12,lambrechts14}. High-resolution images returned by the Philae lander indicate a typical scale closer to 1 cm at the surface \citep{mottola15}, more in agreement with expectations.

Measurements of shape and gravity field have yielded a bulk density of only 0.53 g cm$^{-3}$, so clearly 67P is very porous (70-75\% depending on assumed dust-to-ice ratio). The CONSERT radar \citep{kofman07} had a main aim to measure the internal structure of the comet. Gravity measurements and radar tomography \citep{patzold16,kofman15} indicate that 67P is approximately homogeneous on length scales $<$3 m and very porous. These results are consistent with 67P being a pebble-pile consisting of loosely packed primordial pebbles from the solar protoplanetary disk.

The constituent particles of 67P can also be inferred from the dust particles that fly off the surface. Particles with radii between 2 cm and 1 m have been observed with OSIRIS photometry \citep{rotundi15}. The GIADA instrument \citep{colangeli07} has detected compact (suggesting thermal processing) dust grains of sizes $\leq$100 $\mu$m escaping the comet, but also fluffy, low density ($\rho\sim 10^{-3}$ g cm$^{-3}$) dust aggregates with radii $\sim$0.1-1 mm \citep{rotundi15,fulle15}. The COSIMA instrument \citep{kissel07} collected dust aggregates onto plates to visually analyse their internal structure. The particles collected are porous aggregates that fragment easily upon collision \citep{schulz15}. Low collision speeds ($<$1-10 m s$^{-1}$) and the analysis of the collected aggregates \citep{hilchenbach16} suggest that they are not composed of an ice-dust-mixture and originate from the ice-free surface layers of the comet. \citet{skorov12} presented a comet model consisting of a top layer of ice-free dust aggregates residing on an interior mixture of ice and dust aggregates. Pebble-sized dust aggregates are needed to explain observed comet activity, as the tensile strength of a surface of $\mu$m-sized dust is too high for water sublimation \citep{blum14,blum15}. \citet{gundlach15} applied the model to 67P and found that it can explain the release of observed $\sim$cm-m-sized dust aggregates from the comet surface.

\citet{massironi15} found that 67P is likely a contact binary, inferred from the onion-like structure with shell surfaces centered on the center-of-mass of each separate lobe. Thus 67P may have originally been a binary cometesimal, as is commonly the result of the gravitational collapse model of \citet{nesvorny10}, that later merged gently to a bimodal structure. Altogether, Rosetta and Philae observations of 67P are fully consistent with formation through slow gravitational contraction of a dense cloud of pebbles.

Observations of the comet 103P/Hartley 2 by the EPOXI spacecraft supports the theory of pebble-pile comets. The comet has, like 67P, a bimodal shape and a low density \citep[$\rho\sim$ 0.22-0.88 g cm$^{-3}$ depending on porosity and composition,][]{ahearn11}. EPOXI also found large particles ($\sim$cm-m) in the comet's coma. Investigations by \citet{kretke15}, assuming formation through gravitational collapse, suggest that these particles could be primordial pebbles from which Hartley 2 was formed.

Pluto with its diameter of $\sim$2,400 km is an icy planetesimal on the extreme other end of the size range of Kuiper belt objects. The fly-by by New Horizons showed, surprisingly, that the surface of Pluto is young \citep{stern15}, indicating heating by either short-lived or long-lived radionuclides and recent interior restructuring. Other possible sources of heating (e.g. tidal effects) are, today, insignificant \citep{moore15}. New Horizons is now continuing its journey, through the Kuiper belt, towards the object 2014 MU69, a mid-sized Kuiper belt object (diameter $<$45 km) of the cold population \citep{porter15}. This object has an intermediate size between 67P and Pluto. Its size may be low enough to have avoided extensive particle fragmentation during the collapse \citepalias{wahlberg14}, in contrast to Pluto, and thus maintain its primordial structure the same way as 67P. The results from \citet{lorek16} predict that 2014 MU69 has a dust-to-ice ratio of $\sim$3-7 and constituent pebbles with a volume filling factor close to the maximum value of $\sim$0.4.

\medskip
A major simplification in the work of \citetalias{wahlberg14} was that pebble fragmentation during the collapse was always assumed to be the source of a cloud of $\mu$m-sized monomer particles. In this paper we therefore expand the model for simulating the collapse of pebble clouds with a more realistic fragmentation model. The critical fragmentation speed and fragment size distribution are based on new experimental results presented in a companion paper \citep[hereafter Paper I]{bukhari16}. With this improvement we get more physically correct properties of the resulting planetesimals and can better compare the results with observations of e.g. the next target of New Horizons.

The paper is organized as follows. In \sref{sec:model} we summarize the model and numerical method used in \citetalias{wahlberg14}. The implementation of the results of \citetalias{bukhari16} (the outcome of fragmenting collisions) is described in \sref{sec:fragModel}. The simulations of the collapse of clouds are presented in \sref{sec:sims}. In \sref{sec:disc} we discuss the relevance of our results for the formation of planetesimals by hierarchical accumulation and the validity of neglecting gas drag in our simulations. A discussion of the results and a comparison with previous simulations are presented in \sref{sec:conc}, which is based upon the results of \citetalias{bukhari16}.

\section{Model}\label{sec:model}

A gravitationally bound cloud of pebbles can form e.g. through the streaming instability \citep{youdin05,johansen09,bai10}. In such a cloud, pebbles move around and will eventually collide with each other. The collisions are inelastic, leading to loss of energy and contraction of the cloud. The negative heat capacity property of gravitationally bound systems causes the pebble collision rate (and energy dissipation rate) to increase thanks to the increase in relative speeds and particle density. The result is a runaway collapse, the \textit{gravothermal catastrophe}. This formation process of planetesimals in protoplanetary disks was investigated with numerical simulations in \citetalias{wahlberg14}. In this paper we expand the model to make it more physically realistic. The main change is the model of fragmenting collisions (discussed in \sref{sec:fragModel}).

\subsection{Cloud model}\label{sec:cloud}

We model the pebble cloud in the same way as \citetalias{wahlberg14} with a homogeneous, spherical, non-rotating cloud of initially equal-sized pebbles. By doing this we can treat the cloud as an object characterised by a single single size that strives to get into virial equilibrium at all times. By keeping track of the kinetic and potential energy, the properties (density, collision speeds, free-fall speed, ...) of the cloud can be calculated (knowing the initial values of these properties) with three parameters

\begin{align}
 \eta_\t{eq} &\equiv \frac{E_0}{E}\ , \\
 \eta &\equiv \frac{U_0}{U} = \frac{R}{R_0}\ , \\
 \eta_\t{k} &\equiv \frac{T_0}{T}\ .
\end{align}

\noindent Here $E$ is the total energy, $U$ is the potential energy, $R$ is the radius and $T$ is the kinetic energy of the cloud. The subscript 0 marks the value of the property for the initial cloud. After a collision the values of the parameters change: kinetic energy is dissipated, the cloud contracts, and kinetic energy is released (virialization) thanks to the negative heat capacity. The collapse time of a pebble cloud is short, less than a few orbital periods for planetesimals $\gtrsim$1 km at a Pluto distance from the Sun \citepalias{wahlberg14}. The collapse time decreases with increasing planetesimal size and at some point a size is reached where the energy dissipation is so rapid that the cloud ``wants'' to collapse faster than free-fall. This situation is, of course, not physically possible and arises because it takes some time for the cloud to virialize after a pebble collision. To solve this we add the limitation that the cloud can never contract faster than free-fall. This in turn causes the energy release to slow down so that the pebbles achieve subvirial relative velocities, a situation we refer to as a ``cold'' collapse. Subvirial velocities, in turn, cause lower collision speeds so that a significant fraction of the pebbles will survive the collapse even in massive planetesimals \citepalias{wahlberg14}. As in \citetalias{wahlberg14}, we assume that the individual pebble speeds follow a Maxwellian distribution with the average speed determined by the kinetic energy of the cloud. This is not completely correct since, with dissipative collisions, the pebbles in the cloud do not behave like an ideal gas.

\subsection{Collisional outcomes}\label{sec:collOutcome}

\begin{figure*}[t!]
 \begin{center}
  \resizebox{8.2cm}{!}{\includegraphics{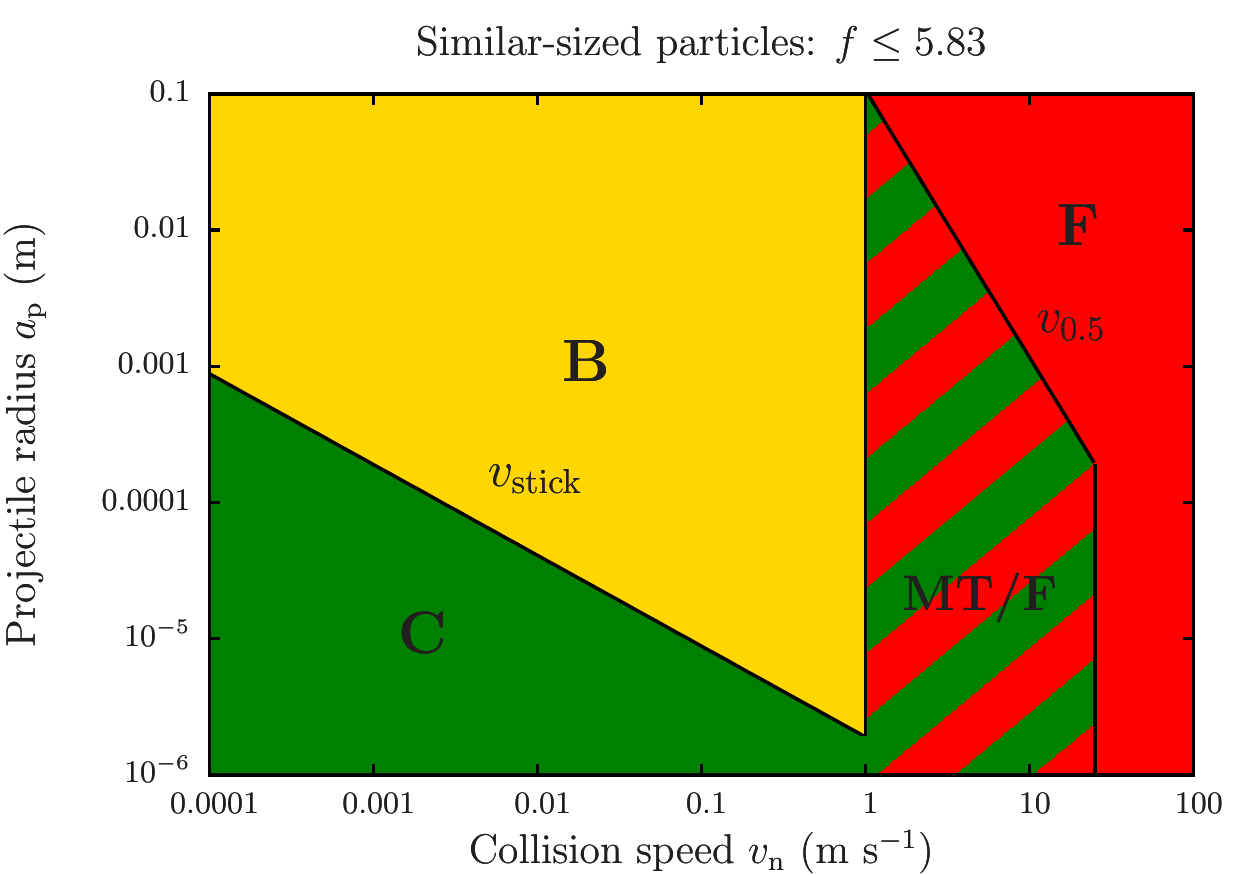}}
  \resizebox{8.2cm}{!}{\includegraphics{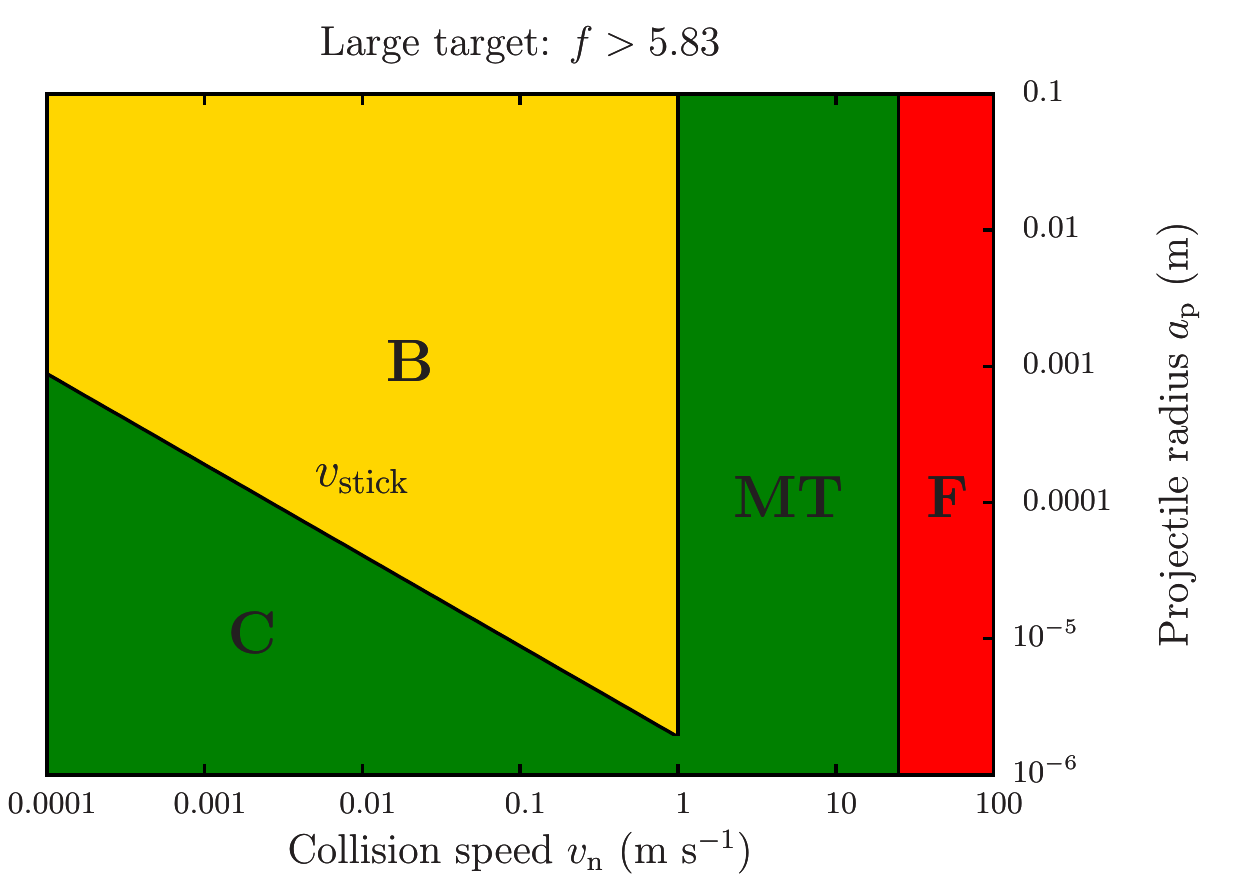}}
  \caption{Schematic map of collisional outcomes for silicate dust aggregates as a function of collision speed, $v_\t{n}$, projectile radius, $a_\t{p}$, and for two regimes of the relative particle size, $f\equiv a_\t{t}/a_\t{p}$ (target radius over projectile radius). Green regions indicate either coagulation or mass transfer (particle growth), yellow regions bouncing, and red regions fragmentation. The dependence of $v_\t{stick}$ on the projectile is described in \eref{eq:vstick}; smaller particles can stick at higher collision speeds. For similar-sized particles ($f\leq 5.83$) and $1\t{ m s}^{-1}\leq v_\t{n}< v_\t{0.5}$ \erefp{eq:v05}, collisions can result in both mass transfer and fragmentation \citepalias[Eq. 2 in][]{bukhari16}. The catastrophic fragmention speed $v_\t{0.5}$ is a function of both target and projectile size \erefp{eq:v05} and cannot be plotted in this two-dimensional map. The curve of $v_\t{0.5}$ shown in the left plot assumes equal-sized particles ($f=1$). The outcome regions are based on \citet{guttler10} and \citetalias{bukhari16}.} \label{fig:outcomeFig}
 \end{center}
\end{figure*}

In our model the cloud collapses through energy loss in inelastic collisions between dust aggregates (pebbles). Depending on particle sizes and collision speed the outcome of a collision can vary. The collision speeds of dust aggregates inside the pebble clouds increase both as the cloud contracts and with increasing planetesimal mass. At a threshold planetesimal size \citepalias[$R_\t{solid}\sim 10$ km,][]{wahlberg14} collisions start to result in fragmentation of the dust aggregates at some point in the collapse. For the result of a collision, in terms of its effect on the target particle, we combine the results from \citetalias{bukhari16} with the results of \citet{guttler10} to include a size distribution of fragments as well as an improved recipe for mass transfer in collisions. Collisions are not necessarily head-on which we correct for by using a randomised impact parameter to calculate the efficient collision speed. When determining the outcome of a collision we use the normal component of the relative velocity, $v_\t{n}$. \tref{tab:outcomeTab} shows the collision outcome for silicate dust aggregates as function of collision speed $v_\t{n}$ and relative particle size $f$ (target radius, $a_\t{t}$, over projectile radius, $a_\t{p}$). For low collision speeds $<$$v_\t{stick}$ the collisions result in coagulation \citep{guttler10}, while higher collision speeds result in either bouncing, mass transfer or fragmentation. The sticking threshold speed can be written as,

\begin{align}
 v_\t{stick} = \sqrt{\frac{5\pi a_0F_\t{roll}}{m_\t{red}}}\ , \label{eq:vstick}
\end{align}

\noindent where $m_\t{red}$ is the reduced mass of the particles, $a_0$ is the monomer radius and $F_\t{roll}$ is the rolling force of the monomers\footnote{We use $F_\t{roll}=(8.5\pm 1.6)\times 10^{-10}$ N for SiO$_2$ spheres with $a_0\sim 1$ $\mu$m from \citet{heim99}.}. The catastrophic fragmentation speed $v_\t{0.5}$, the collision speed required to halve the target, can be written as

\begin{align}
 v_\t{0.5}=\sqrt{2Q^*\left(1+f^3\right)}\ , \label{eq:v05}
\end{align}

\noindent where $Q^*$ is the collision strength \erefp{eq:QStar} and $f=a_\t{t}/a_\t{p}$ is the relative particle size. Collisions are split into two types: two similar-sized particles colliding ($f\leq 5.83$) or a small projectile hitting a large target ($f>5.83$). The difference is that, in the case of a small projectile, mass transfer at higher collision speeds is more likely. In \tref{tab:outcomeTab}, MT/F expresses that collisions in this regime can result in both mass transfer and fragmentation, further discussed in Section 4.1 of \citetalias{bukhari16}. Another difference from the model used in \citetalias{wahlberg14} is the value of the critical size ratio, $f_\t{crit}$. In \citetalias{wahlberg14} a value of $f_\t{crit}=10^{1/3}\approx 2.2$ was used while here we use $f_\t{crit}=5.83$, as advocated in \citetalias{bukhari16}. A schematic map of collision outcomes in $(v_\t{n},a_\t{p})$-space is shown in \fref{fig:outcomeFig}.

\begin{table}
 \begin{center}
  \caption{Outcome regimes of pebble-pebble collisions.} \label{tab:outcomeTab}
  \begin{tabular}{c"c|c}
   $v_\t{n}$ & $f\leq 5.83$ & $f> 5.83$ \\ \thickhline
   $v_\t{n}< v_\t{stick}$ & C & C \\ \hline
   $v_\t{stick}\leq v_\t{n}< 1\t{ m s}^{-1}$ & B & B \\ \hline
   $1\t{ m s}^{-1}\leq v_\t{n}< v_\t{0.5}$ & MT/F & MT \\ \hline
   $v_\t{0.5}\leq v_\t{n}< 25\t{ m s}^{-1}$ & F & MT \\ \hline
   $v_\t{n} \geq 25\t{ m s}^{-1}$ & F & F
  \end{tabular}
  \tablecomments{Here $v_\t{n}$ is the normal component of the relative collision velocity, $f$ is the relative particle size (target radius, $a_\t{t}$, over projectile radius, $a_\t{p}$), $v_\t{stick}$ \erefp{eq:vstick} is the sticking threshold speed \citep{guttler10} and $v_\t{0.5}$ \erefp{eq:v05} is the catastrophic fragmentation speed \citepalias{bukhari16}. For collisional outcomes C denotes coagulation, B bouncing, MT mass transfer and F fragmentation.}
 \end{center} 
\end{table}

\medskip
To get the amount of energy, $\delta E$, dissipated in a collision we use the same equation as in \citetalias{wahlberg14}

\begin{align}
 \delta E=-\frac{1}{2}\mu v_\t{n}^2\left(1-C_\t{R}^2\right)\ , \label{eq:deltaE}
\end{align}

\noindent where $C_\t{R}$ is the coefficient of restitution of the collision. As in \citetalias{wahlberg14} we assume that all kinetic energy in the normal direction of the relative velocity is dissipated, $C_\t{R} = 0$. This is, of course, not completely physical but since it is the square of $C_\t{R}$ that occurs in \eref{eq:deltaE} the value of the coefficient of restitution needs to be relatively high to have a significant effect. \citet{blum93} investigated the coefficient of restitution in collisions between silicate aggregates. The authors split the coefficient of restitution into the normal and the tangential component. From their experiments they find that the normal coefficient of restitution is small ($C_\t{R}^2\sim 0$-0.55). This means that even in grazing collisions, corresponding to high impact parameters, \eref{eq:deltaE} can be used.

\subsection{Dust aggregate mass transfer and fragmentation}\label{sec:fragModel}

The main difference between the simulations in this paper and the ones in \citetalias{wahlberg14} is the treatment of mass transfer and fragmentation. In \citetalias{wahlberg14} mass transfer could only occur in collisions with high mass ratio. Experimental results from \citetalias{bukhari16} show that it can happen for low $f$ as well (row three, column one in \tref{tab:outcomeTab}). Collisions between similar-sized particles in the velocity regime $1\t{ m s}^{-1}\leq v_\t{n}< v_\t{0.5}$ result in either fragmentation or mass transfer. The probability of both survival of the target and mass transfer is approximated as \citepalias{bukhari16}

\begin{align}
p_\t{sur}=\left\{
\begin{array}{lcr}
 0.194f-0.13, &\t{}& 1\leq f\leq5.83 \\
 1, &\t{}& f>5.83.
\end{array}\right.
\end{align}

\noindent More importantly, mass transfer is no longer 100\% efficient but follows the experimental results. The mass transfer efficiency is a function of collision speed and particle sizes and generally lies in the range 10-30\% \citepalias[see][]{bukhari16}.

In the simulations in \citetalias{wahlberg14} we modeled fragmentation as erosion. The collision energy goes into removing monomers from the dust aggregate one by one. This results in a bimodal fragment size distribution with one large remnant and the rest of the mass in monomers (unless the collision energy is enough for complete fragmentation). For this paper, however, we use the results of laboratory experiments of collisions between silicate dust aggregates from \citetalias{bukhari16}, as described below for reference.

From \citetalias{bukhari16} the fragment size distribution can be split into two parts: the largest fragment and a continuous distribution of the remaining fragments. First we need to find the mass of the largest fragment, $m_\t{l}$. \citetalias{bukhari16} finds that the mass of the largest fragment is a function of the collision energy, $E_\t{coll}$, and can be written as a Hill equation

\begin{align}
 \mu \equiv \frac{m_\t{l}}{m_\t{t}} = 1 - \frac{E_\t{coll}^n}{E_{0.5}^n+E_\t{coll}^n}\ , \label{eq:muFrag}
\end{align}

\noindent where $m_\t{t}$ is the target mass, $E_\t{coll}$ is the center-of-mass collision energy, $E_{0.5}$ is the energy required to halve the target (to get a largest fragment of half the mass of the initial target) and $n=0.55$ \citepalias[see][]{bukhari16} is the Hill coefficient. This equation has an s-shape where $E_{0.5}$ marks the region where the curve drops and $n$ describe the steepness of the drop (steeper for higher $n$). For low collision energies $\mu\sim 1$, since the target ``barely'' fragments and for high collision energies most of the target is fragmented, $\mu \sim 0$.

The fragmentation energy $E_{0.5}$ has also been investigated in \citetalias{bukhari16} and found to be dependent on the size of the target and projectile

\begin{align}
 Q^* \equiv \frac{E_{0.5}}{m_\t{t}} = 5.81 \t{ J kg}^{-1}\left(\frac{a_\t{t}}{1\t{ cm}}\right)^{-2.70}\left(\frac{a_\t{p}}{1\t{ cm}}\right)^{1.12}\ , \label{eq:QStar}
\end{align}

\noindent where $Q^*$ is the fragmentation energy divided by the target mass, $a_\t{t}$ the target radius and $a_\t{p}$ the projectile radius.

\begin{figure}
 \begin{center}
  \resizebox{9cm}{!}{\includegraphics{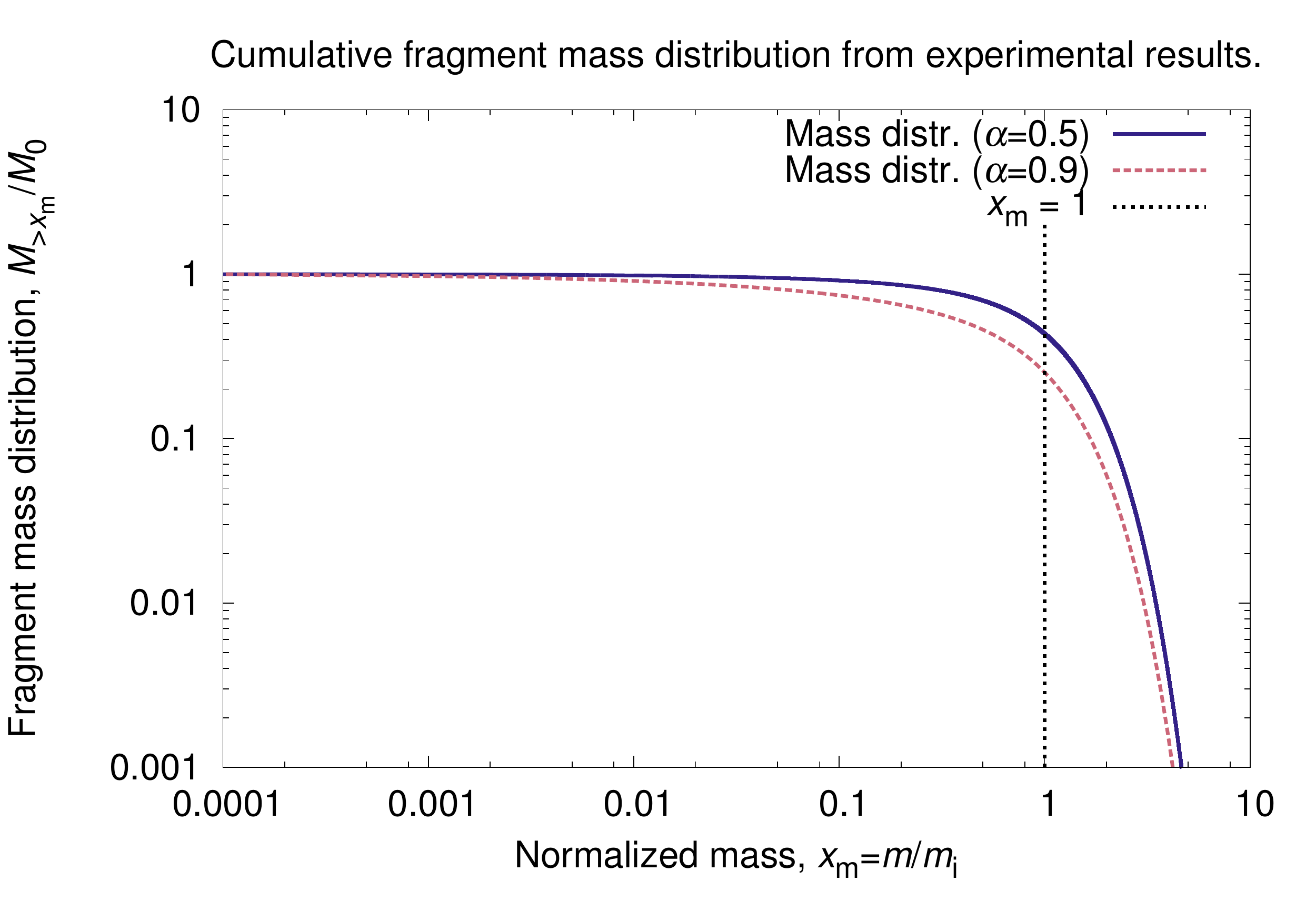}}
  \caption{Cumulative fragment mass distribution after a fragmenting collision in terms of the normalized mass, $x_\t{m}\equiv m/m_\t{i}$. The vertical line shows $x_\t{m}$ for the mass of a fragment with cut-off area $A_\t{i}$ \erefp{eq:cumNum}. A smaller value of the slope $\alpha$ of the fragment size distribution \erefp{eq:cumNum} results in a larger fraction of the total mass in large fragments (red vs. green curve).} \label{fig:fragDistrMod}
 \end{center}
\end{figure}

\citetalias{bukhari16} finds that the cumulative number distribution of the fragments after a fragmenting collision fits very well with a power-law with an exponential cut-off \citepalias[e.g. Fig. 14 in][]{bukhari16}

\begin{align}
 N_{>A} \propto A^{-\alpha} e^{-\left(\frac{A}{A_\t{i}}\right)^\nu}, \label{eq:cumNum}
\end{align}

\noindent where $N_{>A}$ is the number of fragments with a projected area larger than $A$, $\alpha$ the exponent describing the power-law, $A_\t{i}$ the cut-off area and $\nu=2$ an exponent describing the steepness of the exponential cut-off. The value of $\alpha$ is observed to be between $\sim$0.2 and 2 \citepalias[e.g. Fig. 15 in][]{bukhari16} so we do two sets of simulations: one with $\alpha =0.5$ and one with $\alpha =0.9$. In the simulations we are more interested in the distribution of the mass in fragments, not the number, and with \eref{eq:cumNum} we can derive the cumulative mass distribution of the fragments. We use the variable $x_\t{m}\equiv m/m_\t{i}$ where $m_\t{i}$ is the mass of a fragment with projected area $A_\t{i}$ and get

\begin{align}
 \frac{M_{>x_\t{m}}}{M_0} = \frac{1}{I_0}\int^\infty_{x_\t{m}}\left[\alpha x^{-\frac{2}{3}\alpha}+\nu x^{\frac{2}{3}\left(\nu-\alpha\right)}\right]e^{-x^{\frac{2}{3}\nu}}\t{d}x\ , \label{eq:cumMass}
\end{align}

\noindent where $M_{>x_\t{m}}$ is the total mass in fragments with masses larger than $x_\t{m}m_\t{i}$, $M_0$ is the total mass available in fragments and $I_0$ is a normalization constant for the integral. This integral is not analytically solvable and must be solved numerically. We generate arrays with elements on the curve of the solution to use in the simulations. The cut-off area $A_\t{i}$ (and hence $m_\t{i}$ as well) is a function of collision speed and projectile size \citepalias[Eq. 21 and Fig. 16 in][]{bukhari16}. However, by changing variables to $x_\t{m}$ we only need to solve the integral once (for each value of $\alpha$) and can use the same solution throughout all simulations. We then only need to calculate $m_\t{i}$ after each fragmenting collision. \fref{fig:fragDistrMod} shows the solution of \eref{eq:cumMass} for two different values of $\alpha$ (0.5 and 0.9), which are investigated in the cloud collapse process in \sref{sec:sims}. The curves in \fref{fig:fragDistrMod} indicate that large fragments contain most of the total mass after a fragmenting collision. A more shallow slope $\alpha$ in \eref{eq:cumNum} results in even more of the mass in large fragments.

\noindent In the numerical model (a representative particle approach, described in \sref{sec:numeric}) we only need one of the fragments from the mass weighted size distribution \erefp{eq:cumMass}. To select a particle after a fragmenting collision we use a random number, $X$, uniformly distributed between 0 and 1. The largest fragment is not included in \eref{eq:cumMass} and has to be treated separately. We first check if $X\leq \mu\frac{m_t}{m_t+m_p}$ (the fraction of the total mass in the largest fragment); in that case the fragment mass is $\mu m_\t{t}$. Otherwise we solve \eref{eq:cumMass} with our numerical recipe for $x_\t{m}$ using 

\begin{align}
 \frac{M_{>x_\t{m}}}{M_0}=\frac{X-\mu}{1-\mu}, \label{eq:fragMass}
\end{align}

\noindent and select $x_\t{m}m_\t{i}$ as the fragment mass. The random number $X$ in \eref{eq:fragMass}, which is between $\mu$ and 1 in case the largest fragment was not selected, is now renormalized to be uniformly distributed between 0 and 1 in \eref{eq:fragMass}, since \eref{eq:cumMass} is normalized to 1. We additionally impose that the second largest fragment has a maximum allowed mass of $(1-\mu)m_\t{t}$ for $\mu\geq 0.5$ and $\mu m_\t{t}$ for $\mu<0.5$. In case of mass transfer, the cut-off area, $A_\t{i}$, has a different dependence on the collision speed and particle sizes \citepalias[Figs. 16 and 17 in][]{bukhari16}.

%
%

\subsection{Numerical model}\label{sec:numeric}

To be able to follow the collapse process we use, as in \citetalias{wahlberg14}, a representative particle Monte Carlo model \citep{zsom08,ormel07} that uses collision rates to find which particles collide and the time between collisions. The idea with this algorithm is to, out of the large number $N$ of physical particles, randomly select a smaller number, $n$, of representative particles and follow the evolution of them. Each representative particle, $i$, has its own properties (mass, velocity, ...) which can change during the collapse. One can think of a representative particle as one particle in a swarm of identical physical particles. If a property of a representative particle changes, then the property changes for all particles in the swarm. The number of representative particles still needs to be large enough so that the distribution of properties matches the true distribution of properties. Most simulations in this paper are done with $n=250$. The model is described in \citet{zsom08} and the implementation for our simulations in \citetalias{wahlberg14}.

\section{Results}\label{sec:sims}

In our simulations we aim to investigate how the implementation of the fragmentation model from \citetalias{bukhari16} affects the results from \citetalias{wahlberg14}. We are mainly interested in how the size distribution of particles inside the final planetesimal depends on planetesimal mass. The shape of the size distribution will in turn affect the packing efficiency and density of the planetesimal. We are also interested in exploring at what phases of the collapse different collisional outcomes occur. The collision speeds in a massive cloud are high, so fragmenting collisions will take place. However, at some point the energy dissipation is too fast for virialization and the particles will move with subvirial velocities. This means that in the end, collisions result in bouncing or coagulation instead of fragmentation, causing pebbles to survive the entire collapse even for massive planetesimals.

%
%
\begin{figure*}
 \begin{center}
  \resizebox{8.2cm}{!}{\includegraphics{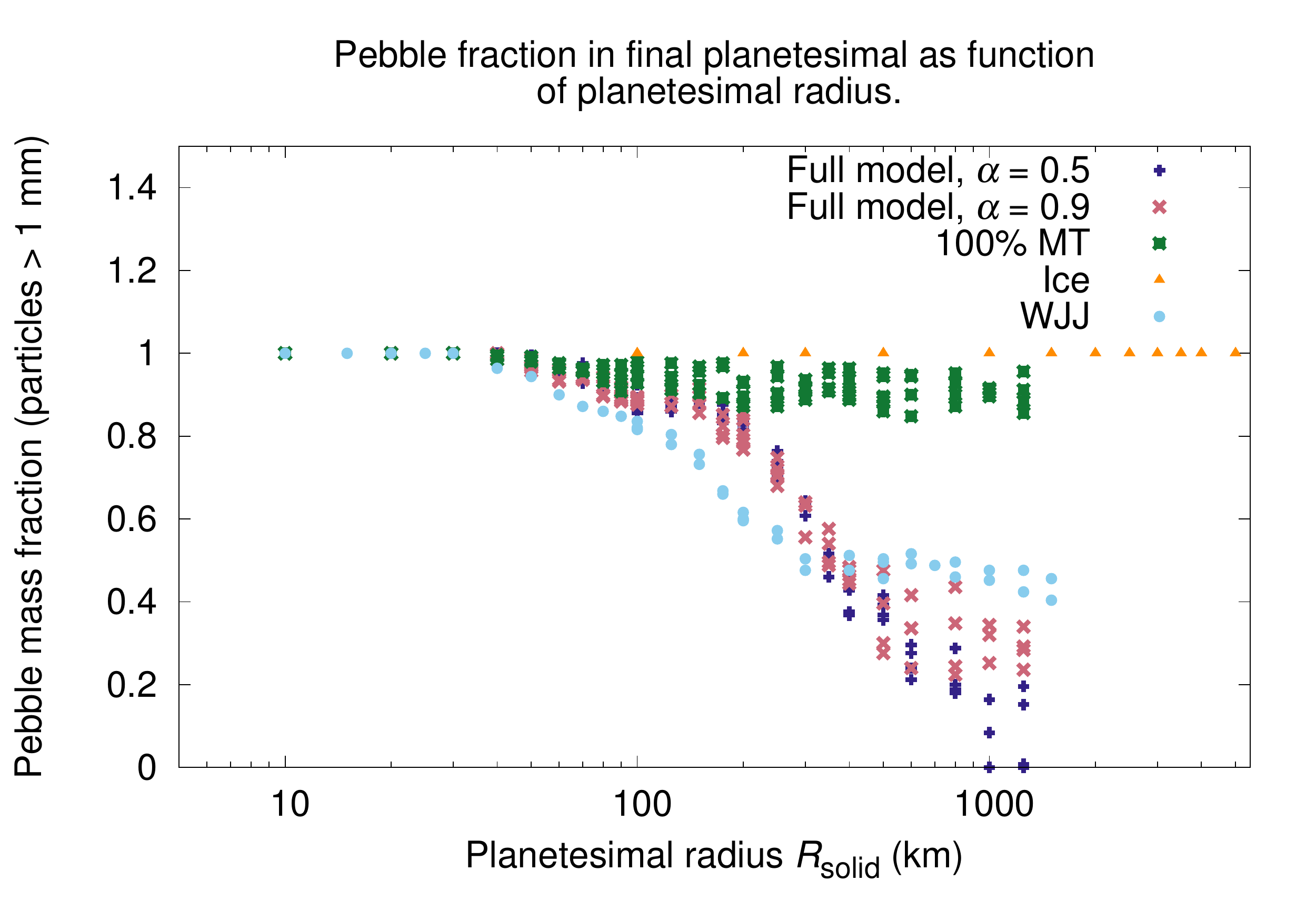}}
  \resizebox{8.2cm}{!}{\includegraphics{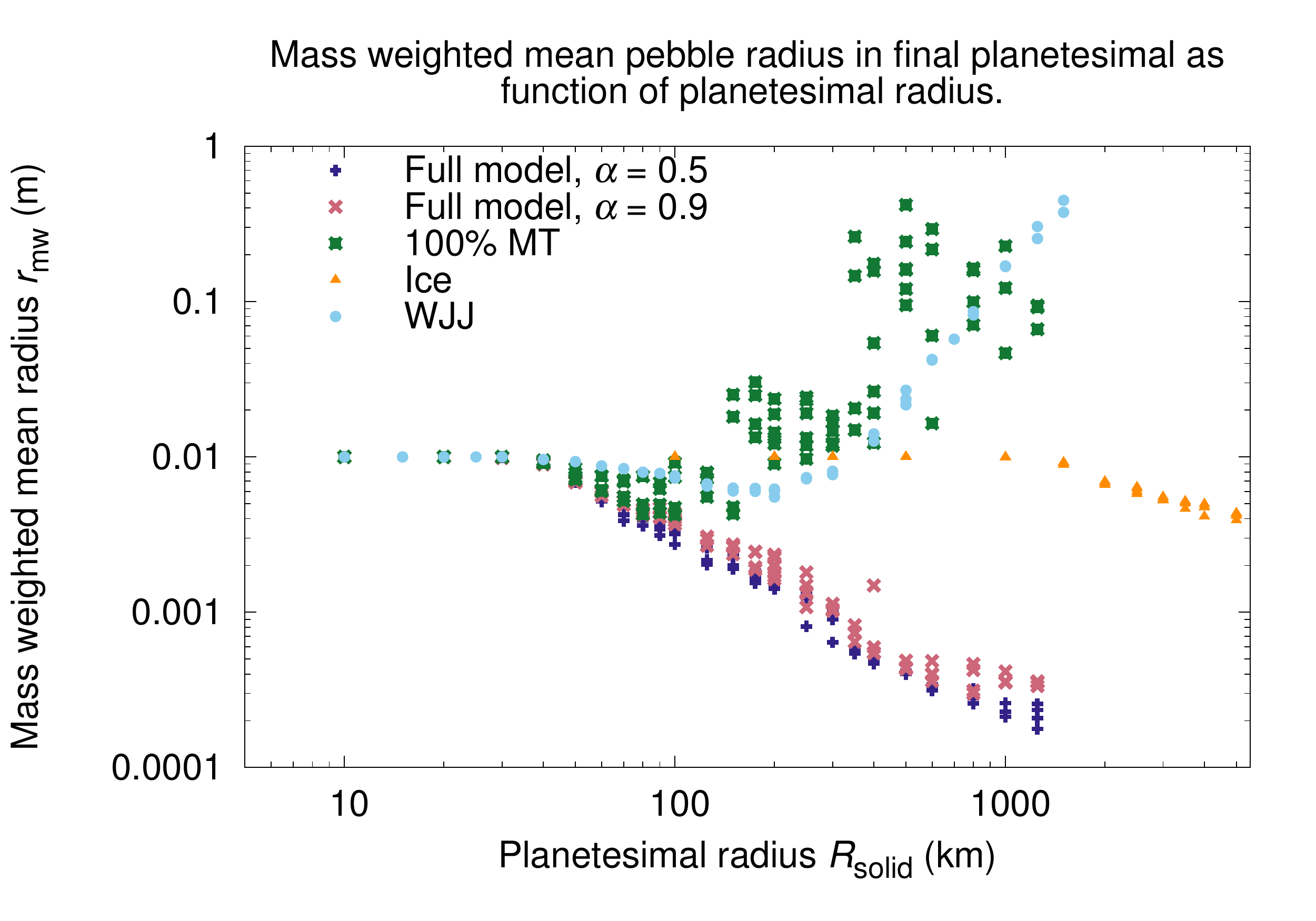}}
  \caption{Fraction of mass in pebbles with radii $>$0.5 mm (left plot) and mass weighted mean radius of particles $\bar{a}_\t{mw}$ (right plot) as function of planetesimal radius for the five sets of simulations. The outcome is very dependent on planetesimal size and model. More massive pebble clouds have higher collision speeds, which result in fragmentation of pebbles. For the full fragmentation model ($\alpha=0.5$ and $\alpha=0.9$) both pebble mass fraction and $\bar{a}_\t{mw}$ drop continuously with planetesimal size, even though the collapse is cold. The value of the power law index of fragments, $\alpha$, does not affect the results very much. For simulations with ice aggregates (\textit{Ice}) the collision speeds required for fragmentation are not reached until $R_\t{solid}\gtrsim$ 1,000 km, which is explained by better sticking capabilities of ice. With 100\% efficient mass transfer (\textit{100\% MT}) the pebble mass fraction stays high even for massive planetesimals and $\bar{a}_\t{mw}$ grows significantly with increasing $R_\t{solid}$. In the collapse many collisions between similar-sized particles occur in the region 1 m s$^{-1}\leq v_\t{n}<v_\t{0.5}$ where mass transfer is a possible outcome. With the \textit{WJJ}-fragmentation model we get the same results as in \citetalias{wahlberg14}: the pebble mass fraction in the final planetesimal drops when fragmentation starts to occur and flattens out for the most massive planetesimals because of the cold collapse. Growth of $\bar{a}_\t{mw}$ for massive planetesimals also occurs for the old fragmentation model thanks to efficient mass transfer.} \label{fig:pebble_frac}
 \end{center}
\end{figure*}

\subsection{Initial conditions}\label{sec:IC}

We run our simulations with the model described in \sref{sec:model} \citepalias[the same algorithm used in][with an updated fragmentation and mass transfer model, \sref{sec:fragModel}]{wahlberg14}. \citetalias{bukhari16} finds a range of values on the slope, $\alpha$, of the fragment size distribution \erefp{eq:cumNum} so we run two sets of simulations ($\alpha=0.5$ and $\alpha=0.9$) to see the effect of $\alpha$ on the result. The density of the dust aggregates is also updated. The experiments in \citetalias{bukhari16} use silicate dust aggregates with a filling factor $\Phi=0.35$, which decreases the density of the pebbles compared to previous simulations in \citetalias{wahlberg14}. This affects the simulations in the sense that planetesimals of the same size now have a smaller mass. Looking at Appendix A in \citetalias{wahlberg14} for the analytic derivation of the collapse time, we see that a smaller filling factor results in a shorter collapse time, 

\begin{align}
 t_\t{coll}=4.1\t{ kyr}\left(\frac{R_\t{solid}}{1\t{ km}}\right)^{-1}\left(\frac{a}{1\t{ cm}}\right)\Phi^{2/3}\left(1-C_R^2\right)^{-1}\ . \label{eq:tCollApprox}
\end{align}

\noindent where $a$ is the radius of the pebbles. Lower mass also causes the collision speeds to be slower, so other results, e.g. the final mass fraction in pebbles, will be different. Otherwise we use the same initial conditions for the simulations as in \citetalias{wahlberg14}, namely a homogeneous, spherical, non-rotating cloud of 1-cm-sized silicate pebbles. The initial size of a cloud is equal to the Hill radius of the mass of the cloud (at a distance from the Sun equal to the semi-major axis of the orbit of Pluto), causing the density (and hence the free-fall time of the cloud) to be independent of the cloud mass. We neglect any effect of surrounding gas on the collapsing cloud (the validity of this assumption is discussed in \sref{sec:gas}).

To further investigate how the fragmentation model affects the outcome of the collapse, we run three more sets of simulations. The first set of these additional simulations uses the new fragmentation model but with 100\% mass transfer efficiency (denoted \textit{100\% MT}). The second set uses the new model but with ice instead of silicate (denoted \textit{Ice}). In this model $v_\t{stick}$ and $v_\t{0.5}$ is increased with a factor 10 to simulate higher sticking capabilities \citep{gundlach15b,lorek16}. Finally, we run the third set of additional simulations with the fragmentation model used in \citetalias{wahlberg14} (denoted \textit{WJJ}) where fragmenting collisions result in erosion and the production of a large remnant and a cloud of $\mu$m-sized dust. In these simulations mass transfer with 100\% efficiency can occur for high mass ratio collisions. The models are summarized in \tref{tab:models}.

\medskip
\begin{table}
 \begin{center}
  \caption{Summary of the models used for the five sets of simulations.} \label{tab:models}
  \begin{tabular}{c"l}
   Model & Description \\ \thickhline
   \textit{Full model} & The model described in \sref{sec:model} with \\
   $(\alpha=0.5)$ & slope $\alpha=0.5$ in the fragment size \\
    & distribution \erefp{eq:cumNum}. \\ \thickhline
   \textit{Full model} & The model described in \sref{sec:model} with \\
   $(\alpha=0.9)$ & slope $\alpha=0.9$ in the fragment size \\
    & distribution \erefp{eq:cumNum}. \\ \thickhline
   \textit{100\% MT} & The model described in \sref{sec:model} ($\alpha=0.9$ in \\
    & \erefnp{eq:cumNum}). Using 100\% mass transfer efficiency \\
    & in the model of \sref{sec:fragModel}. \\ \thickhline
   \textit{Ice} & The model described in \sref{sec:model} ($\alpha=0.9$ in \\
    & \erefnp{eq:cumNum}). Simulating the better sticking \\
    & capabilities of ice compared to silicates by \\
    & increasing $v_\t{stick}$ and $v_\t{0.5}$ with a factor 10 \\
    & \citep{gundlach15b,lorek16}.  \\ \thickhline
   \textit{WJJ} & Use of the fragmentation model in \citetalias{wahlberg14}. \\
    & Fragmenting collisions result in a bimodal \\
    & fragment size distribution: one large remnant \\
    & and a cloud of monomers (erosion). Mass \\
    & transfer is possible in collisions with high \\
    & mass ratio.
  \end{tabular} 
 \end{center}
\end{table}

\subsection{The interiors of planetesimals}

We investigate the formation of planetesimals with solid radii between 10 and 1,000 km (up to a few 1,000 km for simulations with ice particles in order to have any fragmenting collisions at all). We omit simulations of smaller planetesimals in this paper, since the pebble collisions there will only result in bouncing. In our simulations we are interested in cloud collapses with fragmenting pebble-pebble collisions where the updated fragmentation model becomes important. An important result from \citetalias{wahlberg14}, which we find in our new simulations as well, is that, for these planetesimal masses, the collision frequency is so high that the collapse is limited by the free-fall timescale of the pebble cloud. This causes the cloud to collapse cold and the particles inside the cloud to move with subvirial speeds.

\fref{fig:pebble_frac} shows the mass fraction of pebbles (particle radius $>$0.5 mm) in the left plot and the mass weighted mean particle size, $\bar{a}_\t{mw}$, in the right plot for the five models as function of planetesimal radius, $R_\t{solid}$. The plots show that the choice of fragmentation model is very important and affects the outcome significantly. For the full model, both the pebble mass fraction and $\bar{a}_\t{mw}$ decrease with increasing $R_\t{solid}$. However, even massive planetesimals (up to $\sim$Ceres-size) have a significant fraction of their mass in pebbles. Compared to the simulations with the \textit{WJJ}-model, the decrease does not level out in the same way for massive planetesimals. An explanation for this is that with the new fragmentation model the collapse is not as cold as it used to be. The change to larger fragments leads to lower collision rates and less efficient energy dissipation. This way the cloud can get closer to virial equilibrium after each collision. The simulations show that the value of the slope of the fragment size distribution, $\alpha$, does not affect the outcome by a great amount. It is possible that for the most massive planetesimals more pebbles survive for higher $\alpha$. In the case of $\alpha=0.9$ more mass is in small fragments after a fragmenting collision. This makes the energy dissipation more efficient, the collapse colder and collision speeds lower.

%
%
\begin{figure}
 \begin{center}
  \resizebox{9cm}{!}{\includegraphics{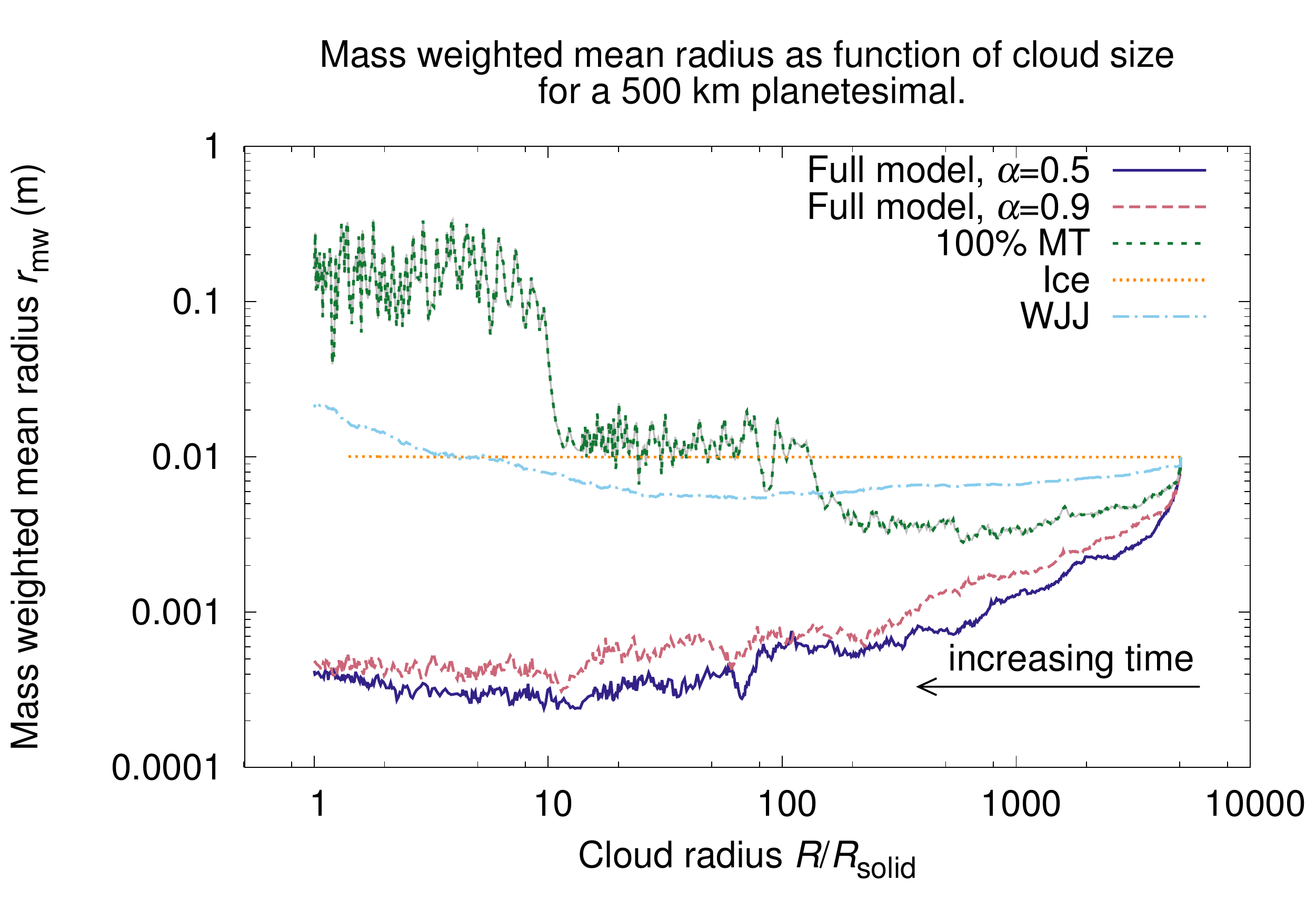}}
  \caption{Evolution of mass weighted mean particle radius $\bar{a}_\t{mw}$ for a $R_\t{solid}=500$ km planetesimal using the different fragmentation models. Note that time increases leftwards in the plot. Fragmentation early in the collapse causes $\bar{a}_\t{mw}$ to drop for all models except the \textit{Ice}-model, since the collision speeds are too low to fragment ice aggregates (\fref{fig:vvir}). Later in the collapse subvirial speeds cause coagulation and mass transfer to start. Efficient mass transfer and relatively high collision speeds (\fref{fig:vvir}) cause substantial growth of particles in the \textit{100\% MT}-model.} \label{fig:rmw_R}
 \end{center}
\end{figure}

Both because of better sticking properties and lower material density of ice, a higher $R_\t{solid}$ is required to have any fragmenting collisions at all when the silicate pebbles are exchanged for ice aggregates (\fref{fig:pebble_frac}). Depending on internal aggregate structure, they can survive collision speeds between $\sim$10 m s$^{-1}$ \citep[compact aggregates,][]{gundlach15b,lorek16} and $\sim$50 m s$^{-1}$ \citep[porous aggregates,][]{wada09}, further discussed in \sref{sec:hokkaido}. The fragmentation in our \textit{Ice}-model simulations (\tref{tab:models}) starts at $\sim$1,000 km and even for a Mars-sized object ($R_\t{solid}\sim$ 3,500 km) all particles in the cloud remain pebbles.

%
%
\begin{figure}
 \begin{center}
  \resizebox{9cm}{!}{\includegraphics{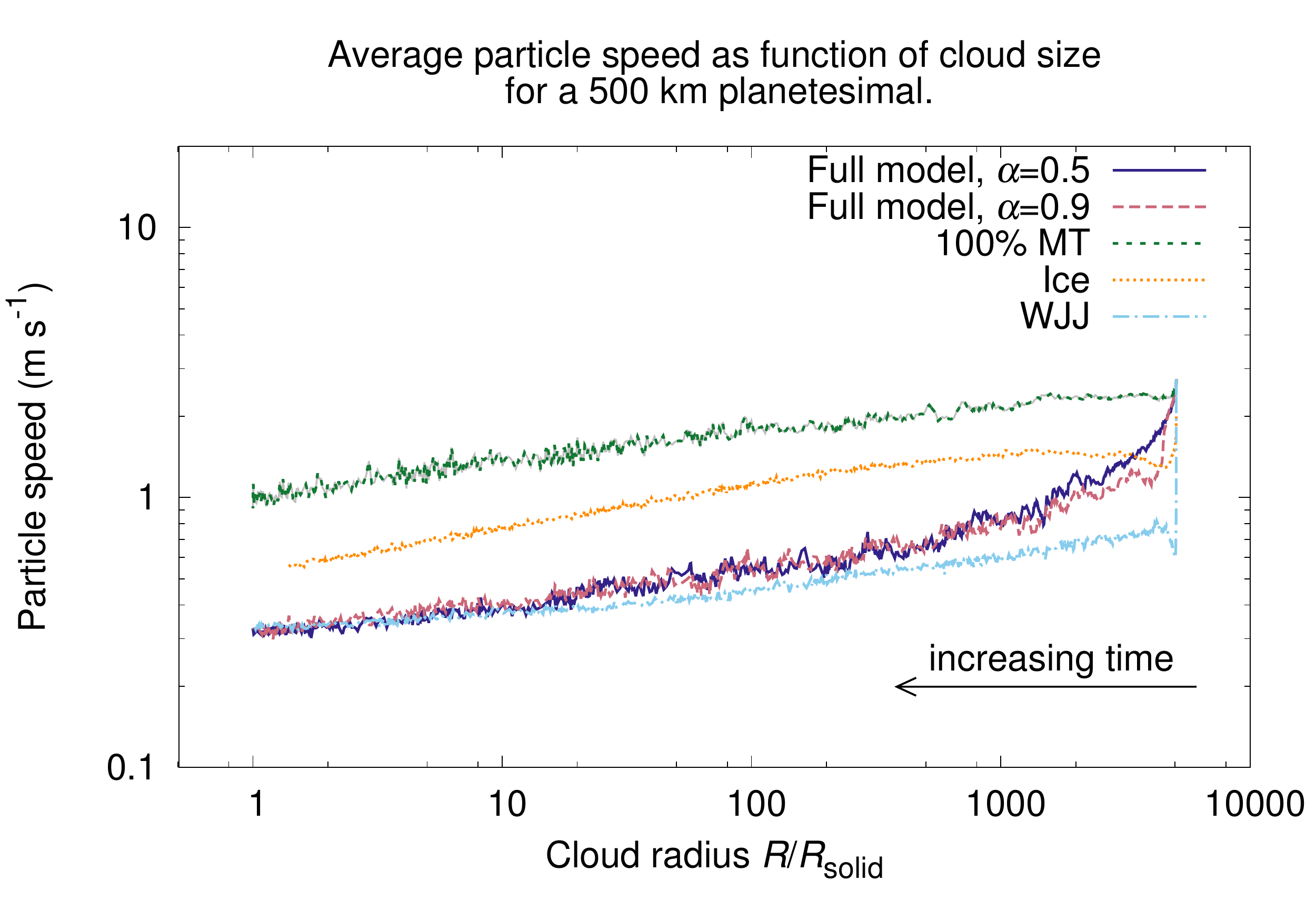}}
  \caption{Evolution of average particle speed (from cloud kinetic energy) for a $R_\t{solid}=500$ km planetesimal using the different fragmentation models. Note that time increases leftwards in the plot. All clouds in this plot are massive with, initially, high particle speeds ($\sim$1 m s$^{-1}$) but their collapse is limited by free-fall so the particle speeds steadily decrease as the collapse continues. Larger particles result in less efficient energy dissipation (lower collision rates) leading to a warmer collapse for the \textit{100\% MT}-model (\fref{fig:sizeDistrFinal}). Collision speeds are dependent on individual particle speeds as well as impact parameter and velocity distribution. This allows fragmenting collisions and decrease of $\bar{a}_\t{mw}$ in \fref{fig:rmw_R} even though curves show speeds $<$1 m s$^{-1}$.} \label{fig:vvir}
 \end{center}
\end{figure}

The only difference between the \textit{100\% MT}-model and the full model is the efficiency of mass transfer. In \textit{100\% MT} all the mass of the projectile particle is transferred while the full model follows the results of \citetalias{bukhari16} (an efficiency of order 10-30\% and the rest of the mass in fragments). The difference in the outcome between the two models is very large. With complete mass transfer, the fraction of mass in pebbles stays around 0.9 even for 1,000 km-sized planetesimals compared to 0-0.4 in the full model. The right plot of \fref{fig:pebble_frac} shows that the pebbles not only survive the collapse but also grow orders of magnitude in size with efficient mass transfer. One should note that, for these particle sizes, the outcome regions from \fref{fig:outcomeFig} might no longer be valid. One reason for the difference is that, in the simulations, a lot of collisions happen in the region 1 m s$^{-1}\leq v_\t{n}<v_\t{0.5}$ \frefp{fig:vvir} where mass transfer is a possible outcome of a collision. In the full model, if mass transfer occurs between two equal-sized particles, only $\sim$55-65\% of the total mass after the collision is in the large particle (10-30\% mass transfer efficiency), while the rest is in small fragments. In the case of full mass transfer all the mass after the collision is in a merged particle. The particle growth is more efficient compared to the \textit{WJJ}-model (where mass transfer also is 100\% efficient), as well. This has several reasons. In the \textit{WJJ}-model, two similar-sized particles cannot transfer mass but only grow in size through sticking at low collision speeds. Mass transfer can occur for higher mass ratios but then the relative mass increase ($\Delta m/m_\t{t}$) per collision is smaller (dust onto a pebble). The collapse process in the \textit{WJJ}-model is also colder, small dust particles dissipate energy faster than larger pebbles. This results in lower particle speeds and mass transfer does not occur to the same degree (the 1 m s$^{-1}\leq v_\t{n}<v_\t{0.5}$-criterion for mass transfer, \fref{fig:outcomeFig}).

%
%
\begin{figure*}
 \begin{center}
  \resizebox{8cm}{!}{\includegraphics{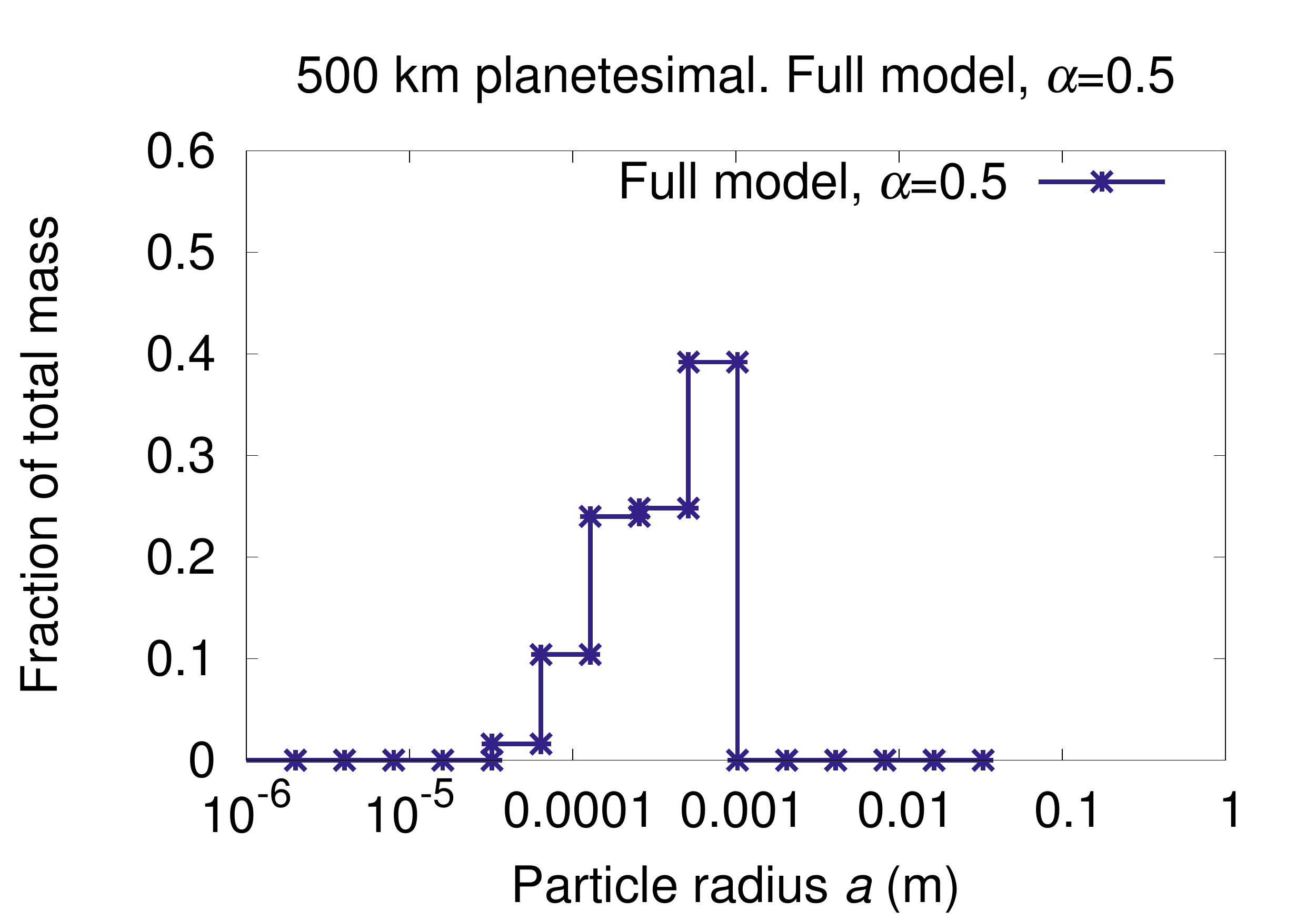}}
  \resizebox{8cm}{!}{\includegraphics{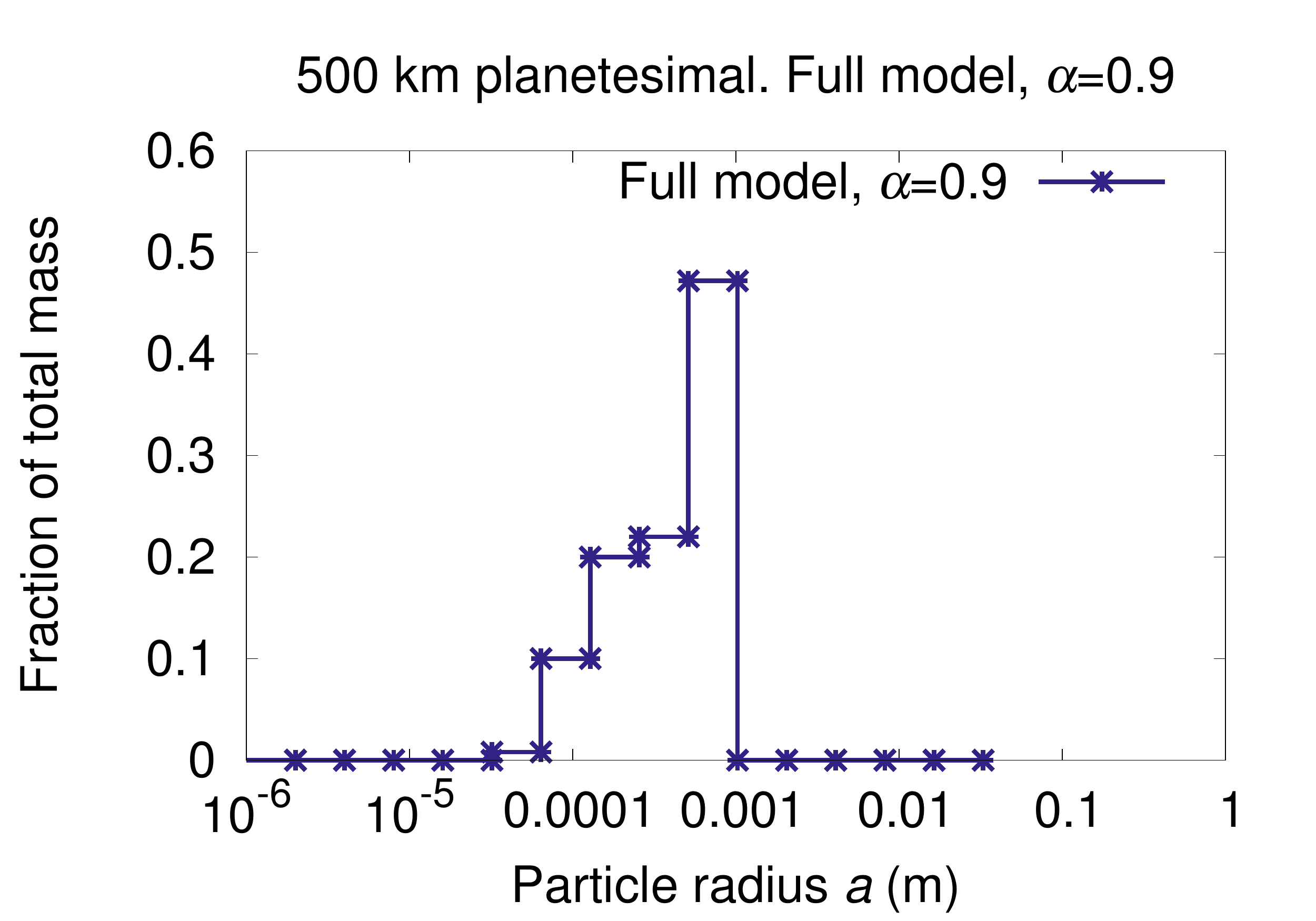}}
  \resizebox{8cm}{!}{\includegraphics{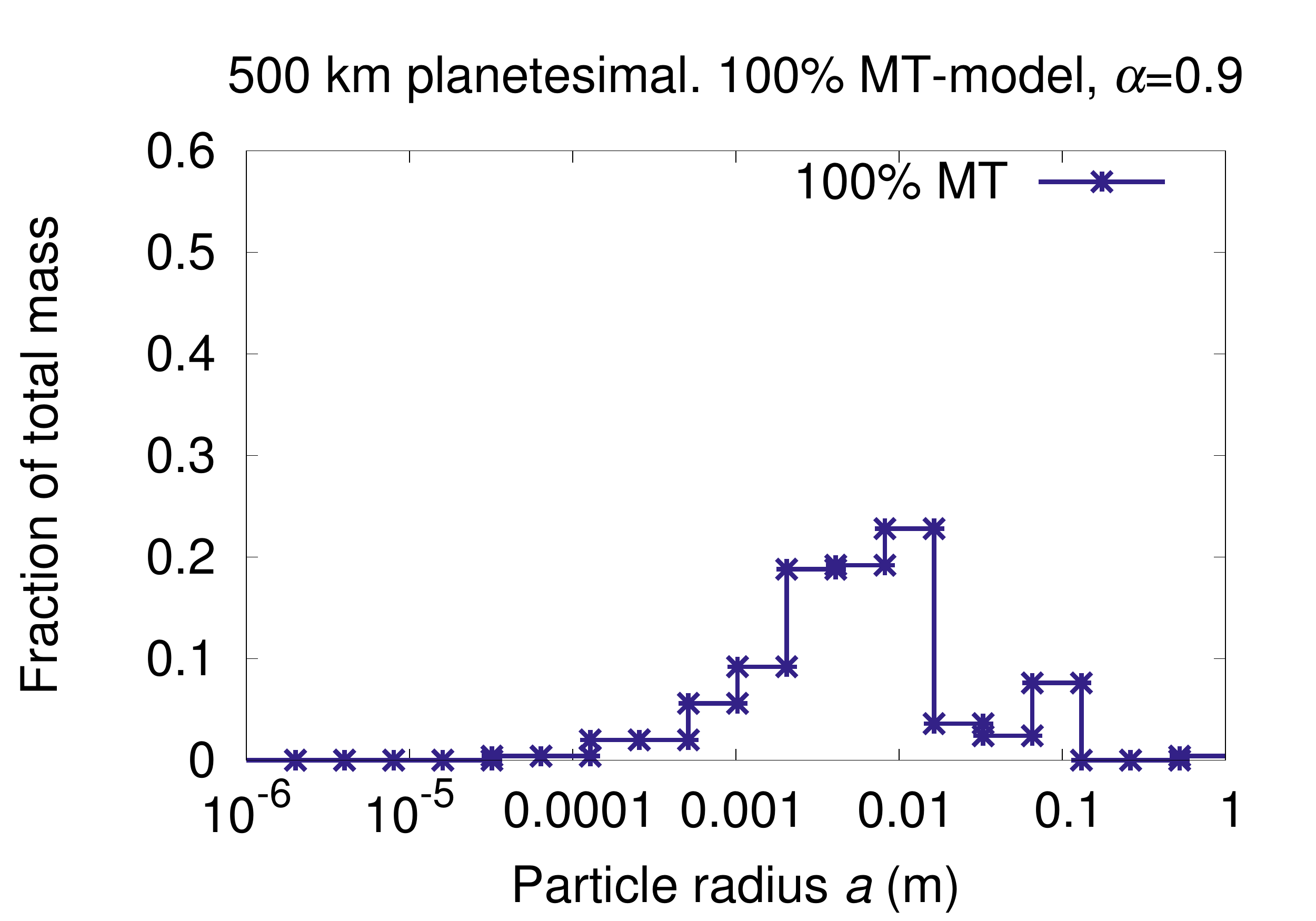}}
  \resizebox{8cm}{!}{\includegraphics{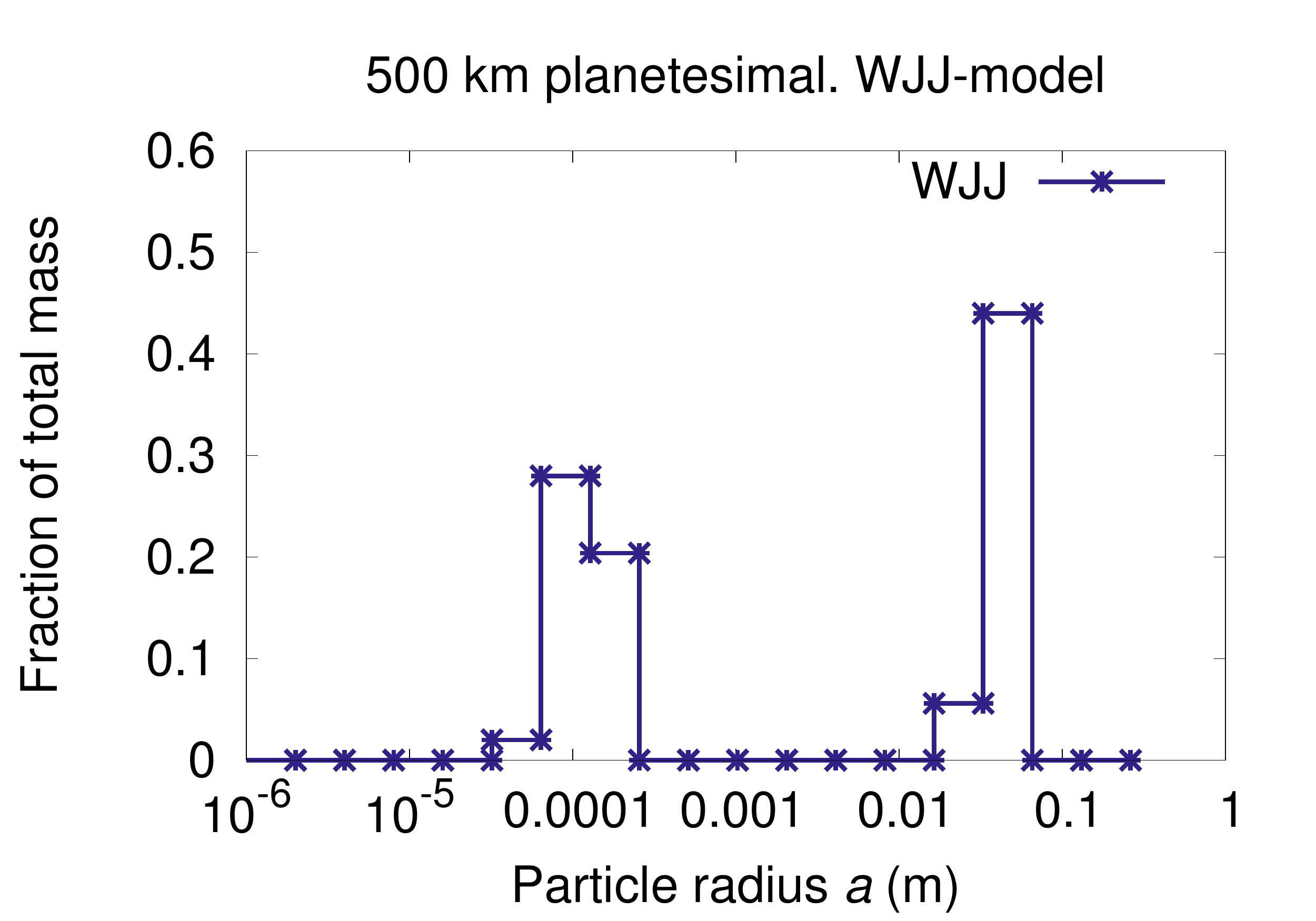}}
  \caption{Mass weighted particle size distribution in a $R_\t{solid}=500$ km planetesimal after collapse using the different fragmentation models described in \tref{tab:models}. All simulations start out with a cloud of cm-sized silicate pebbles. For the full model (top left and top right plots) the pebbles have undergone significant fragmentation and few survive the entire collapse (mm-cm-sized particles). The value of $\alpha$ \erefp{eq:cumNum} does not affect the outcome a great deal. With 100\% mass transfer (bottom left plot), however, the resulting distribution covers a broad range of particle sizes. Collision speeds are in the right range for both fragmentation and mass transfer to occur. The \textit{WJJ}-model (bottom right plot) results in a bimodal particle size distribution: primordial pebbles that have survived the collapse (some growth by mass transfer) and smaller particles formed by coagulation of dust from fragmenting collisions.} \label{fig:sizeDistrFinal}
 \end{center}
\end{figure*}

The energy dissipation efficiency in a collapsing pebble cloud is determined by particle sizes as well as cloud size. \fref{fig:rmw_R} and \fref{fig:vvir} presents the evolution of the mass weighted mean particle radius and average particle speed in the collapse of a $R_\t{solid}=500$ km planetesimal with the different fragmentation models. As in \fref{fig:pebble_frac}, we find that the value of $\alpha$ does not affect the collapse by a great deal. With 100\% mass transfer efficiency, however, the difference becomes significant. By comparing \fref{fig:rmw_R} and \fref{fig:vvir}, one finds that clouds with larger particles (\textit{100\% MT}-model) undergo a warmer collapse (higher collision speeds). Small particles have a higher energy dissipation rate thanks to high collision rates. Even if the average particle speeds in \fref{fig:vvir} are $\lesssim$1 m s$^{-1}$, actual collision speeds can be $>$1 m s$^{-1}$ (we assume a Maxwellian distribution for the particle speeds) leading to fragmentation or mass transfer (\fref{fig:outcomeFig}). This causes $\bar{a}_\t{mw}$ to change over time in \fref{fig:rmw_R}. It increases for the \textit{100\% MT}-model (higher collision speeds and efficient mass transfer) and gradually decreases with time for the full model. The \textit{WJJ}-model initially has rapid fragmentation of a fraction of the pebbles into $\mu$m-sized dust which has an extreme energy dissipation rate and fragmentation rapidly ceases. \fref{fig:sizeDistrFinal} compares the particle size distribution in the resulting 500 km-sized planetesimal for the different fragmentation models. The top two plots show the resulting size distribution for the full model with $\alpha=0.5$ (left) and $\alpha=0.9$ (right). Both models yield a planetesimal consisting of fragmented particles but a significant fraction of the mass in pebbles. Again, the difference between the two models is relatively small. The bottom left plot show the result of the \textit{100\% MT}-model. During the collapse phase of this planetesimal particle collisions resulted both in fragmentation and efficient mass transfer resulting in a size distribution with larger particles than for the full model. The particle size distribution in the planetesimal for the \textit{WJJ}-model is shown in the bottom right plot. In this model, fragmentation is modelled as erosion resulting in a bimodal fragment size distribution. In later stages of the collapse, the dust moves slow enough to stick (\fref{fig:vvir}), grows orders of magnitude in size and there are no $\mu$m-sized monomers in the final planetesimal. Some of the dust also sticks onto the primordial pebbles resulting in growth.

\section{Discussion}\label{sec:disc}

\subsection{Planetesimal formation through hierarchical coagulation}\label{sec:hokkaido}


Gravitational collapse of dense pebble clouds is not the only method to form planetesimals. Another suggested planetesimal formation mechanism to overcome issues with bouncing \citep{zsom10} and fragmentation \citep{brauer08} is hierarchical coagulation. A few lucky particles survive in the protoplanetary disk and grow through mass transfer. \citet{windmark12} investigate the hierarchical coagulation process with numerical simulations of a few cm-sized seed particles in a protoplanetary disk with 100 $\mu$m-sized particles. The authors find that, ignoring the effect of radial drift, 100 m-sized planetesimals can form at a distance of 3 AU from a Sun-like star in $\sim$1 Myr. At a Kuiper Belt distance from the star the timescale would be even longer \citep{johansen14}. Our updated mass transfer and fragmentation algorithms would be highly relevant to further investigate this scenario for planetesimal formation. 

\medskip
In the early stages of planetesimal formation, dust monomers coagulate in low-speed collisions. The low speeds reduce the compactification effect of such collisions and results in fluffy dust aggregates \citep[e.g.][]{okuzumi12}. The high porosity of these aggregates increases the critical collision speed required for fragmentation. N-body simulations of collisions between fluffy ice aggregates with 0.1 $\mu$m-sized monomers show that they survive collision speeds up to $\sim$50 m s$^{-1}$ thanks to efficient energy dissipation and also have efficient mass transfer for high mass ratios \citep{wada09,wada13}. In the case of silicate dust aggregates, the critical velocity, as discussed earlier, is a factor $\sim$10 smaller \citep{gundlach15b,lorek16} but still larger than for similar-sized compact aggregates used in laboratory experiments (left plot in \fref{fig:outcomeFig}, but mind the difference in monomer size). Our numerical experiments with the collapse of clouds of icy pebbles show already that fragmentation is almost absent for all planetesimal sizes. In that case, gravity or internal heating would be needed to compactify the objects \citep[e.g.][]{kataoka13b}.

\subsection{Effect of gas on the collapsing pebble cloud}\label{sec:gas}

In our simulations we ignore the effect of surrounding gas on the collapse. This approximation needs to be validated. To investigate its validity, we can compare the speeds of the aggregates in the cloud with the terminal speed due to gas drag, $v_\t{t}$. The typical speeds of an aggregate in the cloud can be compared to the virial speed, $v_\t{vir},$ or the free-fall speed, $v_\t{ff}$, of the cloud. These can be written as

\begin{align}
 v_\t{t} &= \tau_\t{f}\frac{GM}{R^2}, \label{eq:vter} \\
 v_\t{vir} &= \sqrt{\frac{3GM}{5R}}, \label{eq:vvir} \\
 v_\t{ff} &= \sqrt{\frac{2GM}{R_0}}\sqrt{\frac{R_0-R}{R}}, \label{eq:vff}
\end{align}

\noindent where $\tau_\t{f}$ is the friction time of the pebbles, $M$ is the mass of the planetesimal and $R$ is the radius of the cloud. All clouds start out having Roche density so $R_\t{0}\propto M^{1/3}\propto R_\t{solid}$ which results in the initial value of both terminal and virial speed having the same scaling with planetesimal radius,

\begin{align}
 v_\t{t,0}\propto v_\t{vir,0}\propto R_\t{solid}.
\end{align}

\begin{figure}
 \begin{center}
  \resizebox{9cm}{!}{\includegraphics{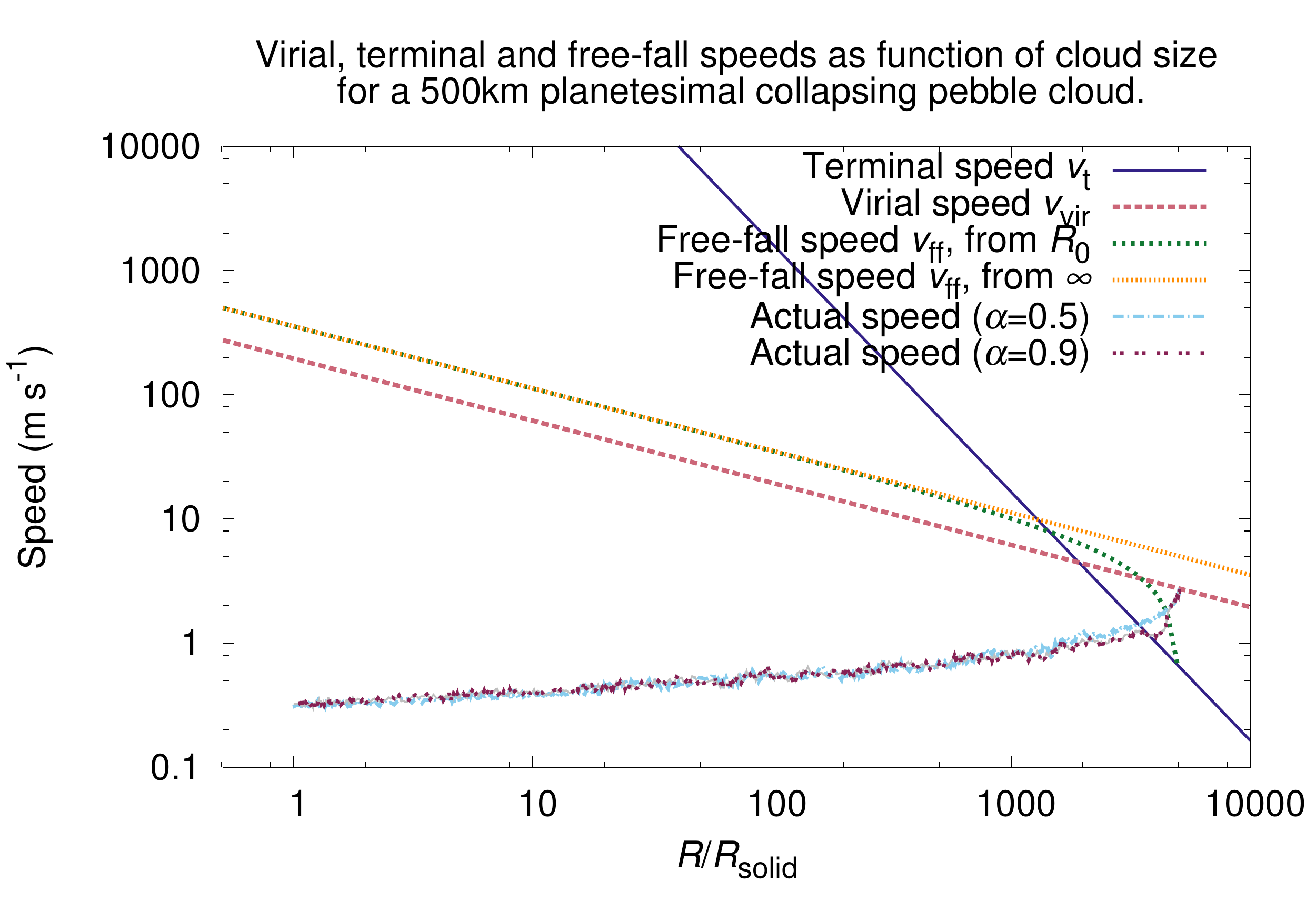}}
  \caption{Comparison between speeds of dust aggregates and the terminal speed, $v_\t{t}$, because of gas drag (Eqs. \ref{eq:vter}-\ref{eq:vff}) for the collapse of a cloud with a solid radius of 500 km. The analytic particle speeds, $v_\t{vir}$ and $v_\t{ff}$ (green, blue and pink curves), are higher than the terminal speed ($v_\t{t}$, red curve) early in the collapse and gas could, very well, affect the collapse. Around cloud size $R\sim$ 1,000-2,000 $R_\t{solid}$ the terminal speed passes the particle speeds. In reality it happens sooner because of the free-fall limited cold collapse (cyan and black curves show the average particle speeds in two simulations).} \label{fig:vels_gas}
 \end{center}
\end{figure}

\noindent Pebbles of cm-sizes in the outer protoplanetary disk experience Epstein drag and we have $\tau_\t{f}\propto a\sim 0.1\Omega^{-1}$, where $a$ is the pebble radius and $\Omega$ is the orbital frequency \citep[see][]{youdin10}. Using \erefCon{eq:vter}{eq:vff} we find that

\begin{align}
 v_\t{t,0} & = 0.646\t{ m s}^{-1} \left(\frac{R_\t{solid}}{500\t{ km}}\right), \\
 v_\t{vir,0} & = 2.75\t{ m s}^{-1} \left(\frac{R_\t{solid}}{500\t{ km}}\right), \\
 v_\t{ff,0} &= 0\t{ m s}^{-1},
\end{align}

\noindent and see that at the start of the collapse $v_\t{t}<v_\t{vir}$ so gas drag will play a role. The different scalings of the speeds with the cloud radius results in an increase in $v_\t{t}$ faster than $v_\t{vir}$ and $v_\t{ff}$ reducing the effect of gas. Using \erefCon{eq:vter}{eq:vff} this transition occurs at $R\sim$ 1,000-2,000 $R_\t{solid}$. In reality it would likely happen sooner the more massive the planetesimal is because the collapse is limited by free-fall causing sub-virial collision speeds (see \fref{fig:vels_gas}). In the collapse of massive pebble clouds, fragmenting collisions will occur. Smaller particles have a lower terminal speed ($\tau_\t{f}\propto a$) and the effect of gas will be larger. We plan to include this effect in a future publication that allows different particle sizes to have different contraction speeds.

\section{Conclusions}\label{sec:conc}

In this paper we have added the results of the silicate dust aggregate collision experiments from \citet[][Paper I]{bukhari16} to the model of planetesimal formation in \citet[][WJJ]{wahlberg14}. The planetesimals are assumed to form by the gravitational collapse of pebble overdensities in protoplanetary disks thanks to energy loss in inelastic collisions between the particles in the cloud. The new model has a more realistic treatment of fragmenting collisions and mass transfer, using the laboratory collision experiments in \citetalias{bukhari16}. To investigate the sensitivity to various aspects of the model, we run three sets of simulations using slightly modified fragmentation models (see \sref{sec:IC}).

As in \citetalias{wahlberg14} the collapse times are short and decrease with increasing planetesimal mass. The collapse speed is, however, limited by free-fall and massive planetesimals ($R_\t{solid}\gtrsim 100$ km) all collapse, roughly, on the free-fall time ($\sim$25 years). The free-fall limit causes massive clouds to undergo a cold collapse where particles move with speeds slower than virial equilibrium speed. Particle speeds decrease by collisional dampening as the collapse progresses (\fref{fig:vvir}), causing pebbles to survive collisions even in massive pebble clouds (\fref{fig:pebble_frac}). In the new model, compared to \citetalias{wahlberg14}, the primordial pebbles have a harder time to survive in the most massive planetesimals. In the left plot of \fref{fig:pebble_frac} we see that, for the full model, a Pluto-sized planetesimal only has $\sim$0-20\% of its mass in pebbles. In the \textit{WJJ}-model the same planetesimal has $\sim$50\% of the mass in particles with radii $\geq$0.5 mm. The main explanation for this is that in the \textit{WJJ}-model fragmentation is modelled as erosion: a fragmenting collision results in a large remnant and cloud of small dust particles. In the new model, however, we have a continuous fragment size distribution \erefp{eq:cumMass} and dust production is rare. Energy dissipation is more efficient with small particles \citepalias[Appendix A in][]{wahlberg14} so with the \textit{WJJ}-model energy is dissipated faster, the collision speeds rapidly become subvirial and fewer fragmenting pebble-pebble collisions occur.

The experiments in \citetalias{bukhari16} find a wide range in steepness, $\alpha$, of the power-law describing the fragment size distribution \erefp{eq:cumNum}. To explore the sensitivity to $\alpha$, we made two sets of simulations for the full fragmentation model: one with a shallow slope (less mass in small fragments, $\alpha=0.5$) and one with a steeper slope (more mass in small fragments, $\alpha=0.9$). The plots in \fref{fig:pebble_frac} show that the difference between the two sets is small compared to the differences to the other fragmentation models.

\fref{fig:pebble_frac} also shows that modelling the mass transfer correctly is important. With efficient mass transfer (\textit{100\% MT}-model) pebbles not only survive the collapse to a larger degree (left plot) but also grow orders of magnitude in size (right plot). The difference between the \textit{100\% MT}-model and the full model is that a collision with mass transfer (MT in \fref{fig:outcomeFig}) results in perfect merger, whereas only 10-30\% of the projectile mass is transferred in the full model. In the collapse, many collisions occur in the velocity regime 1 m s$^{-1}\leq v_\t{n}<v_\t{0.5}$ (\fref{fig:vvir}, \erefnp{eq:v05}) where mass transfer is possible for similar-sized particles (\fref{fig:outcomeFig}). In the full model such collisions result in a slightly larger target but most of the projectile mass in small fragments, while in the \textit{100\% MT}-model the result is one merged target. 

Our results confirm previous suggestions made in \citetalias{wahlberg14}, that low-mass planetesimals (few tens of km or smaller) should consist mainly of primordial pebbles, be very porous and have low internal strength. More massive planetesimals consist of a mixture of pebbles and smaller particles. They would then have a better packing capability and be more dense, in agreement with the size-density correlation observed for Kuiper belt objects \citep{brown13}. Our results use the collision experiments with silicate dust aggregates from \citetalias{bukhari16}, while the outer regions of the Solar System contain a large fraction of ices. Our simulations in which we model the particles as ice change the outcome significantly, since ice particles survive much higher collision speeds. This result may nevertheless change with the inclusion of CO and CO$_2$ ice, which have recently been shown to have equally poor sticking properties as silicates \citep{musiolik16}. Therefore our results obtained with silica particles could be a good proxy for the collapse of actual pebble clouds in the outer regions of protoplanetary disks.

\begin{acknowledgements}
KWJ and AJ were supported by the European Research Council under ERC Starting Grant agreement 278675-PEBBLE2PLANET. AJ was also supported by the Swedish Research Council (grant 2010-3710) and by the Knut and Alice Wallenberg Foundation. KWJ and AJ acknowledges the support from the Royal Physiographic Society in Lund for grants to purchase computer hardware to run the simulations on. MB and JB acknowledge the support through the DFG project SFB 963 ``Astrophysical Flow Instabilities and Turbulence'' in sub-project A07.
\end{acknowledgements}

\bibliographystyle{apj} 
\bibliography{myRef} 


\end{document}